\documentclass[twocolumn,english,prl,floatfix,citeautoscript,nofootinbib,superscriptaddress]{revtex4-2}

\usepackage{amsfonts}
\usepackage{amsbsy}
\usepackage{latexsym,epsfig,graphicx}

\usepackage{dcolumn}
\usepackage{subfigure}
\usepackage{comment}
\usepackage{color}
\usepackage[colorlinks=true, citecolor=blue, urlcolor=blue, linkcolor=blue]{hyperref}
\usepackage{amstext}
\usepackage{amssymb}
\usepackage{setspace}
\usepackage[]{amsmath}
\usepackage{verbatim}

\usepackage{times}

\usepackage{braket}

\begin{document}
\title{Feedback-Induced Dynamical Phases in a Self-Adaptive Quantum Kicked Rotor}

\author{Pan Gao}
\affiliation{Laboratory of Quantum Information, University of Science and Technology of China, Hefei 230026, China}
\affiliation{Anhui Province Key Laboratory of Quantum Network, University of Science and Technology of China, Hefei 230026, China}
\affiliation{CAS Center For Excellence in Quantum Information and Quantum Physics, University of Science and Technology of China, Hefei 230026, China}
\author{Zheng-Wei Zhou}
\affiliation{Laboratory of Quantum Information, University of Science and Technology of China, Hefei 230026, China}
\affiliation{Anhui Province Key Laboratory of Quantum Network, University of Science and Technology of China, Hefei 230026, China}
\affiliation{CAS Center For Excellence in Quantum Information and Quantum Physics, University of Science and Technology of China, Hefei 230026, China}
\affiliation{Hefei National Laboratory, University of Science and Technology of China, Hefei 230026, China}
\affiliation{Anhui Center for Fundamental Sciences in Theoretical Physics, University of Science and Technology of China, Hefei, 230026, China}
\author{Guang-Can Guo}
\affiliation{Laboratory of Quantum Information, University of Science and Technology of China, Hefei 230026, China}
\affiliation{Anhui Province Key Laboratory of Quantum Network, University of Science and Technology of China, Hefei 230026, China}
\affiliation{CAS Center For Excellence in Quantum Information and Quantum Physics, University of Science and Technology of China, Hefei 230026, China}
\affiliation{Hefei National Laboratory, University of Science and Technology of China, Hefei 230026, China}
\author{Xi-Wang Luo}
\email{luoxw@ustc.edu.cn}
\affiliation{Laboratory of Quantum Information, University of Science and Technology of China, Hefei 230026, China}
\affiliation{Anhui Province Key Laboratory of Quantum Network, University of Science and Technology of China, Hefei 230026, China}
\affiliation{CAS Center For Excellence in Quantum Information and Quantum Physics, University of Science and Technology of China, Hefei 230026, China}
\affiliation{Hefei National Laboratory, University of Science and Technology of China, Hefei 230026, China}
\affiliation{Anhui Center for Fundamental Sciences in Theoretical Physics, University of Science and Technology of China, Hefei, 230026, China}

\begin{abstract}
We introduce a self-adaptive Floquet system based on a quantum kicked rotor, in which the kicking strength itself becomes a dynamical variable generated self-consistently through cavity-mediated feedback. A superradiant transition gives rise to cavity-mediated kicking and two competing instability channels, symmetric and antisymmetric, which provide a unified organizing principle for the nonequilibrium Floquet phases. For resonant kicking, their competition produces double-kick dynamics that support resonant ballistic transport and an emergent antiresonance with period-quadrupled rotor evolution, arising from a balance between the two instability channels. Remarkably, for incommensurate kicking, the antisymmetric instability stabilizes a robust period-doubled localized phase with persistent subharmonic dynamics despite the underlying incommensurate driving, revealing localized temporal order absent in conventional kicked rotors. As the feedback strength increases, correlated temporal fluctuations progressively suppress quantum interference, driving crossovers from period-doubled localization to irregular localization and eventually to subdiffusive transport. Our results establish a general framework for self-adaptive quantum-chaotic dynamics and demonstrate how dynamical feedback can fundamentally reshape transport, localization, and temporal order in driven quantum systems.

\end{abstract}

\maketitle


\textit{\textcolor{blue}{Introduction}}.---Understanding nonequilibrium dynamics in driven quantum systems is a central challenge in modern physics~\cite{Bukov2015,Eckardt2017,Weitenberg2021,Mori2023,Sieberer2025}. Among periodically driven systems, the quantum kicked rotor (QKR) serves as a paradigmatic model for exploring quantum chaos, ergodicity, and Floquet dynamics~\cite{Izrailev1990a,Fishman1982,Moore1995,Grempel1984,Manai2015,Santhanam2022}. While its classical counterpart exhibits chaotic momentum diffusion~\cite{Chirikov1979}, the QKR displays fundamentally different behavior due to quantum coherence. In particular, destructive interference suppresses classical diffusion and gives rise to dynamical localization, establishing a profound connection between quantum chaos and Anderson localization in momentum space~\cite{Fishman1982,Grempel1984,Moore1995,Manai2015}. Under special resonance conditions, the QKR further exhibits quantum resonance and antiresonance, characterized by ballistic energy growth and periodic evolution, respectively~\cite{Izrailev1980,Dana1995a,Dana1996a,Poletti2006,Oskay2000,Fishman2002a,Zhang2004b, Ryu2006a,Lepers2008}. 
Recent ultracold-atom experiments have further extended the QKR into the interacting many-body regime, demonstrating how interactions can modify dynamical localization and transport~\cite{Cao2022a,SeeToh2022,Guo2025}. Owing to its rich dynamical behavior, the QKR has become a cornerstone for studying nonequilibrium quantum dynamics~\cite{SeeToh2022,Cao2022a,Sajjad2022,Shimasaki2024,SeeToh2024a,Guo2025,Madani2025a,Olsen2025,Yang2026,Cantillano2026}.

Despite these developments, the kicking protocol in conventional QKR dynamics remains externally prescribed and independent of the state of the system~\cite{Santhanam2022,SeeToh2022,Cao2022a,Sajjad2022,Shimasaki2024,SeeToh2024a,Guo2025,Madani2025a,Olsen2025,Yang2026,Cantillano2026}. A fundamentally different regime emerges when the kicking field itself becomes dynamical and is generated self-consistently by the system it drives. In such a self-adaptive QKR, the kick strength becomes an intrinsic dynamical variable governed by feedback from the driven system, potentially reshaping transport, localization, and stability. 
This raises two fundamental questions: Can such a self-adaptive QKR be realized in a physical system? If so, how does mutual coupling between the drive and the driven system reshape nonequilibrium quantum dynamics?

In this work, we address these questions by proposing an experimentally feasible self-adaptive QKR using an ultracold Bose gas in an optical cavity, where cavity-mediated feedback turns the kicking strength into a self-consistent dynamical variable. Cavity-mediated dynamical feedback has enabled a variety of collective self-organization phenomena~\cite{Nagy2010a,Baumann2010c,Nagy2011,Baumann2011,Klinder2015,Ferri2021,Domokos2002a,Nagy2008a,PhysRevLett.108.043003,Leonard2017,PhysRevLett.120.263202,Landini2018a,Kroeze2018,Zupancic2019,Helson2023,Piazza2015e,Kessler2019,Chiacchio2019a,Kessler2021,Dreon2022a,Kongkhambut2022,Gao2023,Cosme2025}. We find that this feedback-controlled kicking gives rise to a rich set of Floquet dynamical phases beyond those of conventional QKRs. 
Specifically, an analytical instability analysis reveals two competing superradiant channels, symmetric and antisymmetric, which quantitatively predict the transition thresholds and organize the nonequilibrium Floquet phases.
For resonant kicking, their competition produces double-kick dynamics supporting both resonant ballistic transport and an emergent antiresonance with period-quadrupled rotor evolution, the latter arising from a balance between the two channels. Remarkably, for incommensurate kicking, the antisymmetric instability stabilizes a robust period-doubled localized phase with persistent subharmonic dynamics, revealing localized temporal order absent in conventional QKRs. As the feedback strength increases, correlated temporal fluctuations progressively suppress quantum interference, driving crossovers toward irregular localization and eventually to subdiffusive transport. Our results establish a general framework for self-adaptive quantum-chaotic dynamics and demonstrate how feedback can fundamentally reshape transport, localization, and temporal order in driven quantum systems.

\begin{figure}[t]
\includegraphics[width=1.0\linewidth]{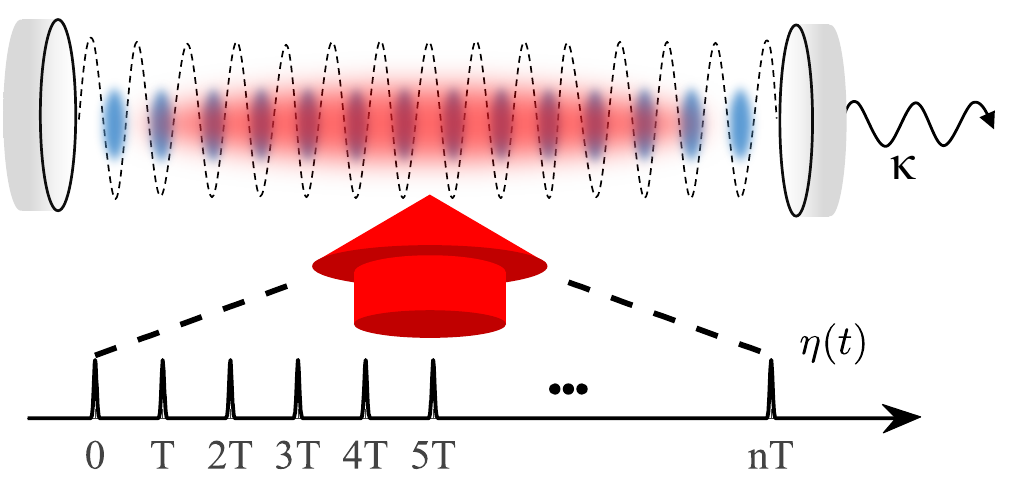}
\caption{Schematic of the proposed setup. A quasi one-dimensional ultracold Bose gas trapped inside an optical cavity is driven by a periodically pulsed transverse pump field $\eta(t)$.}
\label{fig:setup}
\end{figure}

\textcolor{blue}{\textit{Model}}.---We consider an optical cavity coupled to a quasi one-dimensional ultracold Bose gas that is driven by a periodically pulsed transverse pump, as illustrated in Fig.~\ref{fig:setup}. 
The pump and cavity fields are far detuned from atomic transition and coupled through two-photon Raman scatterings~\cite{Nagy2010a,Baumann2010c,Ritsch2013c,Mivehvar2021}.
The pulsed pumping generates cavity-assisted momentum kicks for the atoms, while the intracavity field is simultaneously determined by the atomic density distribution, giving rise to a self-consistent feedback mechanism between atomic dynamics and cavity photons.
Within the mean-field approximation, the system is described by the normalized atomic condensate $\psi(x,t)$ and coherent cavity field $\alpha(t)$, respectively. Introducing the dimensionless variables $\theta=k_c x$ and $\tau=t/T$ (with $k_c$ and $T$ the kicking momentum and period, respectively), the coupled atom-cavity dynamics obey~\cite{SUPP}
\begin{eqnarray}
i\partial_{\tau} \psi(\theta,\tau)
&=&
\left[
-\frac{\hbar_s}{2}\partial_{\theta}^2
+\sum_{n\in \mathbb{Z}} \delta(\tau-n)
 K\cos\theta
\right]
\psi, \label{eom1} \\
  i\partial_{\tau}\alpha(\tau)
&=&
\left(
-\Delta_c-i\kappa
\right)
\alpha
-
\sum_{n\in \mathbb{Z}} \delta(\tau-n)\,
\eta
\Theta.
  \label{eom2}
\end{eqnarray}
Here, $\hbar_s=2\omega_RT$ denotes the dimensionless effective Planck constant with $\omega_R=\hbar k_c^2/2m$ the recoil frequency, $\Delta_c$ and $\kappa$ characterize the cavity detuning and dissipation, respectively. 
The periodically pulsed cavity-assisted potential realizes an effective quantum kicked rotor with a dynamically generated kick strength $K(\tau)=2\eta\,\mathrm{Re}[\alpha(\tau)]$
with $\eta$ the effective transverse driving strength. Unlike conventional QKRs with externally prescribed kick strength, here $K$ is determined self-adaptively.  
The cavity field provides a nonlinear feedback channel that dynamically connects the kick strength to the instantaneous atomic density order $\Theta(\tau)=-\int_0^{2\pi} d\theta |\psi(\theta,\tau)|^2\cos\theta$.
To highlight the role of cavity feedback, we focus on the noninteracting Bose gas and neglect the weak cavity-induced ac-Stark shift, whose effects are discussed in~\cite{SUPP}. Unless otherwise stated, the system is initialized in a homogeneous condensate with an infinitesimal cavity-field fluctuation, and the nonequilibrium phases are identified from their long-time dynamics.

\begin{figure}[t]
\includegraphics[width=\linewidth]{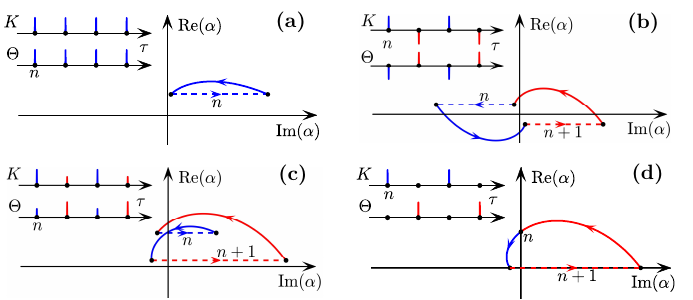}
    \caption{Schematic of the cavity-field evolution associated with different instability channels and their competition.
(a) Symmetric instability channel.
(b) Antisymmetric instability channel.
(c) Mixed response with a dominant symmetric component, giving rise to resonant transport.
(d) Balanced competition between the two channels, leading to feedback-induced antiresonance.
The cavity-field trajectory in the complex plane is shown together with the corresponding stroboscopic kick strength and density order parameter.
Solid (dashed) lines denote free evolution (kick) processes.}
\label{fig:instability}
\end{figure}

\textcolor{blue}{\textit{Superradiance and instability mechanism}}.---Since the cavity-mediated kicking strength is proportional to the intracavity field, nontrivial kicked-rotor dynamics can only emerge when a finite cavity field is generated. We therefore begin by analyzing the stability of the normal phase, where the atoms remain in a stable state and the cavity field vanishes [i.e.,
$\psi(\theta,\tau)=\psi_s,
\alpha(\tau)=0$].
The evolution over one driving period is described by the Floquet map
\begin{eqnarray}
    \psi(\theta,n_-+1)&=&e^{i\hbar_s\partial_{\theta}^2/2}e^{-iK(n_-)\cos{\theta}}\psi(\theta,n_-)  \label{eq:Floquet1}\\
    \alpha(n_-+1)&=&e^{i\Delta_c-\kappa} \left[\alpha(n_-)+i\eta\Theta(n_-)\right],
    \label{eq:Floquet2}
\end{eqnarray}
where $n_-$ denotes the time immediately before the $n$th kick. During a kick, the cavity field is displaced according to $\alpha\rightarrow\alpha+i\eta\Theta$, while the condensate acquires a phase imprint $\psi\rightarrow e^{-iK\cos\theta}\psi$. Here $K$ represents the cavity-induced kick strength while $\Theta$ characterizes the atomic density modulation that drives the cavity field.

To determine the onset of superradiance, we consider small fluctuations around the normal-state solution
$\delta\alpha(\tau),
\delta\Theta(\tau)$
where the induced kicking fluctuation is given by
$\delta K(\tau)=2\eta\mathrm{Re}[\delta\alpha(\tau)].$
Linearizing Eqs.~(\ref{eq:Floquet1}) and (\ref{eq:Floquet2}) yields a Floquet stability matrix that maps the fluctuations over one driving period,
and we seek Floquet eigenmodes satisfying
\begin{eqnarray}
    [\delta\alpha(n_-+1),\delta\Theta(n_-+1)]
=\lambda
[\delta\alpha(n_-),\delta\Theta(n_-)],
\label{eq:Floquet_eig}
\end{eqnarray}
where $\lambda$ is the Floquet multiplier~\cite{1998}. Since only the real part of the cavity field enters the feedback loop and the density order parameter $\Theta$ is real, the relevant instability channels are characterized by real Floquet multipliers. The normal phase is stable when $|\lambda|<1$, while $|\lambda|>1$ signals an instability toward a superradiant state. The transition boundary is therefore determined by $|\lambda|=1$.
From Eqs.~(\ref{eq:Floquet1})-(\ref{eq:Floquet_eig}), one obtains
$\delta\alpha(n_-)=\eta\cdot \chi_c(\Delta_c,\kappa,\lambda)\cdot\delta\Theta(n_-)$ and $\delta\Theta(n_-)=\eta\cdot \chi_a(\psi_s,\hbar_s,\lambda)\cdot\mathrm{Re}[\delta\alpha(n_-)]$, from which the critical pump strength is determined as 
$\eta_c(\lambda)=1/\sqrt{\chi_a\cdot\mathrm{Re}[\chi_c}]$.
The expression of $\chi_a$ and $\chi_c$ are given in~\cite{SUPP}.

Remarkably, two instability channels emerge, corresponding to a symmetric one with $\lambda=1$ and an antisymmetric one with $\lambda=-1$. 
Generally, $\eta_c(\pm1)$ are different but may be close to each other. 
Representative Floquet trajectories at the instability threshold are shown in Figs.~\ref{fig:instability}(a) and \ref{fig:instability}(b), where the symmetric instability recovers after one driving period, whereas the antisymmetric instability changes sign after one period and recovers only after two periods, giving rise to a subharmonic response. 
Above threshold, these two channels naturally favor period-one and period-doubled cavity responses, respectively. 
Their competition underlies the emergent nonequilibrium phases of the self-adaptive QKR.

\begin{figure}[t]
\includegraphics[width=\linewidth]{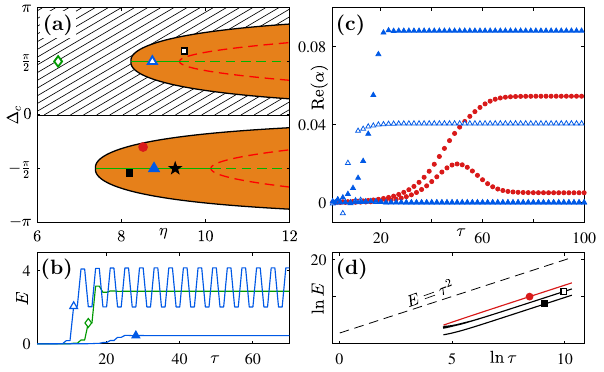}
\caption{Resonant kicking dynamics for $\hbar_s=\pi$ and $\kappa=4$.
(a) Dynamical phase diagram in the $\eta$-$\Delta_c$ plane, showing the uniform (white region) and Bessel dark (shaded region) phases below threshold, together with the resonant transport (orange regions) and feedback-induced antiresonance (green lines). Beyond the red dashed lines, the dynamics become increasingly sensitive to initial fluctuations.
(b) Evolution of the atomic energy in the Bessel dark phase (diamond markers) and the antiresonant phase (triangle markers). The antiresonance generally exhibits period-quadrupled atomic dynamics, which reduce to period-doubled for $\Delta_c<0$ and $\eta$ below the star.
(c) Stroboscopic evolution of the cavity field in the resonant (circles) and antiresonant (triangles) phases.
(d) Evolution of the atomic energy for several resonant cases (different markers), all exhibiting ballistic transport with $E\propto\tau^2$.
The parameters used in (b)-(d) correspond to the marked points in panel (a).
}
\label{fig:phase_diagram1}
\end{figure}

\textcolor{blue}{\textit{Resonant Kicking Dynamics}}.---We first consider the resonant regime and focus on
$\hbar_s=\pi$.
After folding the free-evolution phase 
into the Floquet Brillouin zone~\cite{Santhanam2022},
momentum states with even and odd indices acquire only two distinct kinetic phases.
In the conventional QKR,
this resonance condition leads to ballistic momentum transport and quadratic energy growth~\cite{Ryu2006a,Santhanam2022}.
We find that, cavity-mediated feedback qualitatively enriches this behavior and generates additional dynamical phases absent in the standard QKR model.

The resulting dynamical phase diagram in the $\eta$-$\Delta_c$ plane is shown in Fig.~\ref{fig:phase_diagram1}(a).
At weak pumping below the superradiant threshold,
the cavity field vanishes, and
the Bose gas occupies either the uniform zero-momentum state or a Bessel dark state, depending on the cavity detuning.
The uniform condensate remains stable for red detuning ($\Delta_c<0$), but 
becomes unstable and evolves into
a Bessel dark state 
$\psi_s(\theta,n_-)\propto e^{-iA\cos\theta}$
for blue detuning ($\Delta_c>0$),
with energy evolution shown in Fig.~\ref{fig:phase_diagram1}(b) and $A$ determined by 
$J_1(2A)=0$~\cite{SUPP}.
The superradiant phase boundaries obtained from the instability analysis agree well with the numerical simulations.
Above threshold,
the cavity field becomes finite and generates self-adaptive kicking.
The resulting dynamics are controlled by the interplay between the symmetric and antisymmetric instability channels discussed above.

Far away from
$|\Delta_c|=\frac{\pi}{2}$,
one instability channel dominates (the symmetric channel for $|\Delta_c|<\frac{\pi}{2}$ and the antisymmetric for $|\Delta_c|>\frac{\pi}{2}$).
Although the cavity field exhibits a period-doubled response with alternating kick amplitudes $K_1$ and $K_2$, as illustrated schematically in Fig.~\ref{fig:instability}(c) and verified numerically in Fig.~\ref{fig:phase_diagram1}(c),
a Floquet transfer-matrix analysis 
reveals that resonance is preserved over two driving periods~\cite{SUPP}.
Consequently,
the atoms exhibit ballistic momentum transport and quadratic energy growth $E(\tau)\propto\tau^2$, as confirmed numerically in Fig.~\ref{fig:phase_diagram1}(d).
Note that, near the threshold, the superradiance may be governed by a single instability channel with $K_2=\pm K_1$, which also leads to resonance.

A qualitatively different behavior emerges at
$|\Delta_c|=\frac{\pi}{2}$,
where the symmetric and antisymmetric instability channels become comparable.
In this regime,
every second kick is dynamically suppressed,
corresponding to
$K_2=0$.
The system therefore experiences an effective kick only every two driving periods,
which maps the dynamics onto an effective kicked rotor with
$\hbar_s^{\rm eff}=2\hbar_s=2\pi$.
As a consequence,
ballistic transport is replaced by periodic evolution,
giving rise to an emergent feedback-induced antiresonance even under resonant driving.
The microscopic mechanism can be understood from the cavity-field evolution shown in Fig.~\ref{fig:phase_diagram1}(c) and schematized in Fig.~\ref{fig:instability}(d). At
$|\Delta_c|=\pi/2$,
free cavity evolution rotates the cavity field by $\pi/2$ in the complex plane during one driving period.
As a result,
a finite real cavity field responsible for atomic kicking is transformed into an imaginary field after free evolution,
suppressing the subsequent kick.
Meanwhile,
the atomic density modulation generated by the previous kick pumps the cavity field and restores its real component during the following cycle.
This feedback loop produces a kick sequence self-organized into
$K_1,0,K_1,0,\cdots$,
thereby dynamically generating the antiresonance condition.
Generally, the condensate recovers its initial state only after two effective kick cycles~\cite{Dana1995a,Dana1996a,Poletti2006}, resulting in period-quadrupled dynamics [see Fig.~\ref{fig:phase_diagram1}(b)]. We also find a period-doubled antiresonant regime ($\eta<\eta_\star$) for red detuning, in which the atomic state recovers after one effective kick cycle~\cite{SUPP}.
The balance between the two instability channels is progressively lifted as $\Delta_c$ moves away from $\pm\frac{\pi}{2}$.
The suppressed kick regains a finite amplitude,
and the feedback-induced antiresonant behavior continuously crosses over into ballistic resonant transport.

In the strong-driving regime, the dynamics exhibits fluctuations and becomes sensitive to the initial state~\cite{SUPP}, yet the competition between symmetric and antisymmetric instability channels continues to organize the Floquet dynamics and sustain robust resonant and antiresonant responses.
The interplay between symmetric and antisymmetric instabilities thus provides a unified mechanism for both ballistic transport and feedback-induced antiresonance, the latter emerging spontaneously from cavity-atom self-adaption rather than externally imposed driving.

\begin{figure}[t]
\includegraphics[width=\linewidth]{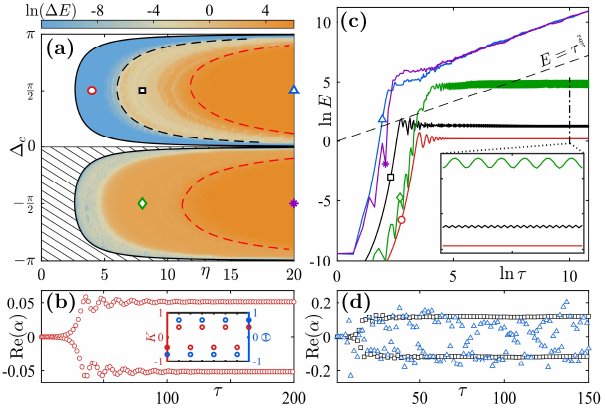}
\caption{Incommensurate kicking dynamics for $\hbar_s=6.8$ and $\kappa=4$.
(a) Dynamical phase diagram in the $\eta$-$\Delta_c$ plane. White and shaded regions denote the uniform and finite-momentum dark phases, respectively. The color scale in superradiant phases shows the energy broadening over the interval $\tau\in[8\cdot10^4, 10^5]$, revealing crossovers from period-doubled localization to irregular localization and eventually to subdiffusion. The black and red dashed lines ($\ln\Delta E=-5$ and $4$) approximately separate these regimes.
(b) Stroboscopic evolution of the cavity field in the period-doubled localized phase. Inset: the corresponding kick strength and density order parameter.
(c) Condensate energy for representative parameter sets (different markers), exhibiting bounded oscillations in the localized regimes and subdiffusive growth with approximately $E\propto\tau^{2/3}$. Inset: enlarged view of the bounded oscillations.
(d) Stroboscopic evolution of the cavity field in the irregular localized and subdiffusive regimes. Parameters used in (b)-(d) are marked in panel (a).
}
\label{fig:phase_diagram2}
\end{figure}

\textcolor{blue}{\textit{Incommensurate Kicking Dynamics}}.---We now turn to the incommensurate regime and take $\hbar_s=6.8$ as a representative example. In a conventional quantum kicked rotor, incommensurate kicking gives rise to dynamical localization through quantum interference~\cite{Santhanam2022}. The key question here is therefore how cavity-mediated feedback reshapes localization physics. 
As we show below, weak feedback preserves localization, whereas stronger feedback generates correlated temporal fluctuations that gradually destroy localization and give rise to subdiffusive transport. Meanwhile, the antisymmetric instability channel supports a robust period-doubled localized phase, revealing a new form of localized subharmonic order.

The phase diagram in the
$\eta$--$\Delta_c$ plane is shown in Fig.~\ref{fig:phase_diagram2}(a).
For blue-detuned driving ($\Delta_c>0$), the superradiant transition is governed by the antisymmetric instability channel. Below threshold, the atoms remain in the zero-momentum state, corresponding to a local maximum of the folded Floquet dispersion. Above threshold, the cavity field develops a robust period-doubled response, as shown in Fig.~\ref{fig:phase_diagram2}(b). Although the underlying Floquet dynamics is incommensurate,
cavity-mediated feedback locks the system onto the antisymmetric instability channel and suppresses the irregular temporal fluctuations. 
Consequently, localization coexists with a stable subharmonic response, producing a feedback-stabilized period-doubled localized regime. This behavior sharply contrasts  with the conventional incommensurate QKR, whose localized dynamics remain irregular and quantum-chaotic~\cite{Santhanam2022}.
The self-organized kicking field drives the atoms into a momentum distribution in which the complex relative phases accumulated during the incommensurate free evolution are compensated by the subsequent cavity-assisted kick. Meanwhile, odd momentum states acquire an additional $\pi$ phase, thereby reversing the sign of the density order ($\Theta\rightarrow-\Theta$), which in turn pumps the cavity field into the opposite configuration ($\alpha\rightarrow-\alpha$).
After two driving periods, both $\Theta$ and $\alpha$ recover their original values, producing a robust subharmonic response. During this evolution, the momentum-space population distribution and hence the kinetic energy $E$ remains unchanged [see Fig.~\ref{fig:phase_diagram2}(c)]. 
The robust subharmonic dynamics here emerge from the self-consistent interplay between cavity feedback and the antisymmetric instability channel. 

As the pump strength increases, cavity-mediated feedback becomes progressively stronger, where the incommensurate dynamical phases can no longer be compensated. The cavity field develops increasingly irregular temporal fluctuations, which act as an intrinsic dynamical noise source for the kicked rotor and gradually destroy the period-doubled dynamics. The corresponding energy and cavity field evolutions are shown in Figs.~\ref{fig:phase_diagram2}(c) and \ref{fig:phase_diagram2}(d), respectively. For moderate feedback strengths, these fluctuations remain weak and correlated, leading to irregular localization.  However, once the fluctuation amplitude becomes sufficiently large, the temporal correlations of the kick sequence gradually decay and the effective kick strength behaves more similarly to that of a noisy kicked rotor~\cite{Cohen1991d,Klappauf1998b,Ammann1998a,Steck2000,Sarkar2017}. Quantum interference is then progressively suppressed, leading to subdiffusive momentum transport with an approximate power-law growth close to $E(\tau)\propto\tau^{2/3}$. This anomalous energy growth indicates that the feedback-induced fluctuations retain temporal correlations~\cite{SUPP} rather than acting as purely random external noise~\cite{Ammann1998a,Klappauf1998b,Steck2000}.
The system therefore exhibits a crossover from a dynamically localized regime to a feedback-induced subdiffusive regime.

For red-detuned driving ($\Delta_c<0$), the symmetric instability dominates. Below the superradiant threshold, the homogeneous condensate evolves into a momentum-selected dark state $\psi_s\propto\cos(6\theta)$,
corresponding to a minimum of the folded Floquet dispersion. Above threshold, cavity-mediated kicking emerges and the system enters a dynamically localized regime, as shown in Fig.~\ref{fig:phase_diagram2}(a). In contrast to the blue-detuned case, no robust period-doubled dynamics are observed because the antisymmetric instability superradiance is absent. Nevertheless, the overall localization-to-diffusion crossover remains qualitatively unchanged: weak feedback suppresses temporal fluctuations and preserves localization, whereas strong feedback generates increasingly irregular cavity dynamics and eventually induces subdiffusive energy growth with $E\sim \tau^{2/3}$, as shown in Fig.~\ref{fig:phase_diagram2}(c).

\textcolor{blue}{\textit{Conclusion}}.---In summary, we have investigated a self-adaptive QKR realized in a cavity-QED system under periodically pulsed transverse driving, where the kicking strength becomes a dynamical variable generated self-consistently through cavity-mediated feedback. We find that two competing superradiant instability channels, symmetric and antisymmetric, provide a unified organizing principle for the nonequilibrium dynamical phases, giving rise to feedback-induced antiresonance, localized subharmonic order, and anomalous transport. These results demonstrate that making the kicking field dynamically responsive to the driven system can qualitatively reshape quantum-chaotic dynamics, leading to dynamical phases beyond those of conventionally driven QKRs. Importantly, the parameter regimes supporting these novel dynamical phases are experimentally accessible~\cite{SUPP}, making the predicted phenomena readily testable with current cavity-QED platforms. More broadly, our work establishes a general framework for self-adaptive quantum-chaotic dynamics and provides a route toward feedback-controlled transport, localization, and temporal order in driven quantum systems. 

The present framework can be further extended to incorporate strong interactions and quantum fluctuations beyond mean-field theory. Together with generalizations to higher-dimensional kicked-rotor models, these extensions may reveal new localization phenomena and emergent phases arising from the interplay between many-body correlations and dynamical feedback.

\textcolor{blue}{\textit{Acknowledgments}.---}This work is supported by the National Natural Science Foundation of China (Grants No. 12574544 and No. 12474366) and Quantum Science and Technology-National Science and Technology Major Project (Grant No. 2021ZD0301200).

\textcolor{blue}{\textit{Data availability.---}}The data that support the findings of this article are publicly available~\cite{data}.

%

\begin{widetext}

\section*{Supplemental Material}

\subsection{A. Dynamical equations and experimental parameter estimations}
\label{app:dynamical equations}
We consider a quasi one-dimensional ultracold Bose gas coupled to an optical cavity and driven by a periodically pulsed transverse pump. 
Both the pump and cavity fields are far detuned from the atomic transition, and they are coupled through a two-photon Raman scattering process. We restrict the motion of the atoms
along the cavity axis $x$ by assuming additional tight trapping in the other directions. In the rotating frame with respect to the pump frequency, the Hamiltonian of the system reads~\cite{Mivehvar2021}
\begin{equation}
    H=-\hbar\Delta_c \hat{a}^\dagger \hat{a}+ \int dx \hat\Psi^\dagger(x) \left[\frac{p^2}{2m}+\hbar U_0 \hat{a}^\dagger \hat{a} \cos^2(k_cx)+\hbar\eta(t) (\hat{a}^\dagger+\hat{a})\cos(k_cx)+\frac{\hbar g_{\rm a}}{2}\hat\Psi^\dagger(x)\hat\Psi(x)\right]\hat\Psi(x) + \text{const.},
\end{equation}
where $\hat{a}$ and $\hat\Psi$ are the field operators of the cavity and the Bose gas, $k_c$ is the wave number of the cavity mode. The $\Delta_c$ term in the Hamiltonian corresponds to cavity detuning, the $U_0$ term corresponds to cavity-field-induced ac-Stark shift, while the $\eta(t)$ term describes the two-photon Raman scattering process 
associated with the pump and cavity photon and it is determined by the transverse pump strength. The $g_{\rm a}$ term characterizes the atom-atom interaction. For the periodically pulsed pump, we have 
\begin{equation}
    \eta(t)=\eta_0\sum_n \delta (t-nT)
\end{equation}
with $T$ the driving period and $n$ an integer.

Within the mean-field approximation, the ultracold Bose gas and cavity field are described by the condensate wavefunction $\braket{\hat\Psi}=\psi(x,t)$ and coherent cavity-field amplitude $\braket {\hat{a}}= \alpha(t)$, respectively. The dynamical equations of motion can be obtained from the Heisenberg equations, which lead to~\cite{Mivehvar2021}
\begin{equation}
\begin{aligned}
i\frac{\partial \psi}{\partial t}
&=
\bigg[
-\frac{\hbar}{2m}\frac{\partial^2}{\partial x^2}
+ g_{\rm a}|\psi|^2
+ U_0 |\alpha|^2\cos^2(k_cx)
+2 \eta(t)\operatorname{Re}[\alpha]\cos(k_cx)
\bigg]\psi, \\
i\frac{\partial \alpha}{\partial t}
&=
\bigg[
-\Delta_c-i\kappa
+U_0
\int dx\,|\psi(x,t)|^2\cos^2(k_cx)
\bigg]\alpha
+
\eta(t)
\int dx\,|\psi(x,t)|^2\cos(k_cx).
\end{aligned}
\label{eq:appA_dimensional_cavity}
\end{equation}
We have included the cavity loss $\kappa$ in the equation, and the atomic wave function is normalized to the total atom number $\int dx\,|\psi(x,t)|^2=N$.

We now introduce the dimensionless parameters
\begin{eqnarray}
    \theta=k_cx,\quad
    \tau=\frac{t}{T},\quad
    \tilde\Delta_c=\Delta_cT,\quad
    \tilde\kappa=\kappa T, \quad \tilde U_0= U_0T,\quad \tilde\eta(\tau)=\eta(t) T=\eta_0\sum_n \delta(\tau-n).   
\end{eqnarray}
Then the mean-field wavefunction becomes $\psi(x,t)\rightarrow \psi(\theta,\tau)$ and $\alpha(t)\rightarrow \alpha(\tau)$ with
dynamical equation
\begin{equation}
\begin{aligned}
i\frac{\partial \psi(\theta,\tau)}{\partial \tau}
&=
\bigg[
-\frac{T\hbar k_c^2}{2m}\frac{\partial^2}{\partial \theta^2}
+  g_{\rm a}T  |\psi|^2
+ \tilde U_0 |\alpha|^2\cos^2(\theta)
+2 \tilde\eta(\tau)\operatorname{Re}[\alpha]\cos(\theta)
\bigg]\psi, \\
i\frac{\partial \alpha}{\partial \tau}
&=
\bigg[
-\tilde\Delta_c-i\tilde\kappa
+\tilde U_0
\int d\frac{\theta}{k_c} \,|\psi|^2\cos^2(\theta)
\bigg]\alpha
+
\tilde\eta(\tau)
\int d\frac{\theta}{k_c}\,|\psi|^2\cos(\theta).
\end{aligned}
\end{equation}
By defining the 
recoil frequency and the effective Planck constant
\begin{equation}
    \omega_R=\frac{\hbar k_c^2}{2m},  \quad \hbar_s=2\omega_RT.
\end{equation}
we re-normalize the 
field wavefunction as
\begin{eqnarray}
\tilde\psi(\theta,\tau)=\sqrt{\frac{L}{2\pi N}}\psi(x,t), \quad \tilde\alpha=\frac{1}{\sqrt{N}}\alpha
\end{eqnarray}
with $L$ the size of the atomic system.
The equations of motion reduce to
\begin{equation}
\begin{aligned}
i\frac{\partial \tilde\psi(\theta,\tau)}{\partial \tau}
&=
\bigg[
-\frac{\hbar_s}{2}\frac{\partial^2}{\partial \theta^2}
+  \tilde g_{\rm a}  |\tilde\psi|^2
+ \tilde U_0 {N}|\tilde\alpha|^2\cos^2(\theta)
+2 \tilde\eta(\tau)\sqrt{N}\operatorname{Re}[\tilde\alpha]\cos(\theta)
\bigg]\tilde\psi, \\
i\frac{\partial \tilde\alpha}{\partial \tau}
&=
\bigg[
-\tilde\Delta_c-i\tilde\kappa
+\tilde U_0{N}
\int_0^{2\pi} d \theta \,|\tilde\psi|^2\cos^2(\theta)
\bigg]\tilde\alpha
+
\tilde\eta(\tau)\sqrt{N}
\int_0^{2\pi} d\theta\,|\tilde\psi|^2\cos(\theta).
\end{aligned}
\end{equation}
where $\tilde g_{\rm a}=2\pi T g_{\rm a}\frac{ N}{L}$ is the dimensionless atom-atom interaction strength, and normalization of the wavefunction is 
\begin{eqnarray}
\int_0^{2\pi}|\tilde\psi(\theta,\tau)|^2d\theta=1,
\end{eqnarray}
where we have assumed that the condensate wavefunction has a period of $2\pi$ along $\theta$. After absorbing the $\sqrt{N}$ enhancement factor into $\eta=\sqrt{N}\eta_0$ and $U={N}U_0$, and dropping the tilde for simplicity, we obtain the dynamical equations
\begin{equation}
\begin{aligned}
i\frac{\partial \psi(\theta,\tau)}{\partial \tau}
&=
\bigg[
-\frac{\hbar_s}{2}\frac{\partial^2}{\partial \theta^2}
+  g_{\rm a}  |\psi|^2
+ U |\alpha|^2\cos^2(\theta)
+2 \eta\operatorname{Re}[\alpha]\cos(\theta)\sum_n\delta(\tau-n)
\bigg]\psi, \\
i\frac{\partial \alpha}{\partial \tau}
&=
\bigg[
-\Delta_c-i\kappa
+ U
\int_0^{2\pi} d \theta \,|\psi|^2\cos^2(\theta)
\bigg]\alpha
+
\eta\sum_n\delta(\tau-n)
\int_0^{2\pi} d\theta\,|\psi|^2\cos(\theta).
\end{aligned}
\end{equation}
To highlight the role of cavity feedback, 
in the main text, we have focused on the noninteracting limit with $g_{\rm a}=0$ and neglected the weak cavity-induced ac-Stark shift by setting $U=0$. The atom-atom interaction can be tuned to zero via a Feshbach resonance~\cite{Chin2010,Petrucciani2026}, while the cavity-induced ac-Stark shift can also be made negligibly weak by choosing proper photonic polarization and atomic detuning~\cite{Meng2023}. Nevertheless, their effects
will be discussed later.
Under the conditions $U=g_{\rm a}=0$, we arrive at the dynamical equation used in the main text
\begin{eqnarray}
i\partial_{\tau} \psi(\theta,\tau)
&=&
\left[
-\frac{\hbar_s}{2}\partial_{\theta}^2
+\sum_{n\in \mathbb{Z}} \delta(\tau-n)
 K\cos\theta
\right]
\psi, \label{eq:eom1S} \\
  i\partial_{\tau}\alpha(\tau)
&=&
\left(
-\Delta_c-i\kappa
\right)
\alpha
-
\sum_{n\in \mathbb{Z}} \delta(\tau-n)\,
\eta
\Theta.
  \label{eq:eom2S}
\end{eqnarray}
The dynamically generated kick strength is $K(\tau)=2\eta\,\mathrm{Re}[\alpha(\tau)]$
with $\eta$ the effective transverse driving strength. The parameter $\Theta(\tau)=-\int_0^{2\pi} d\theta |\psi(\theta,\tau)|^2\cos\theta$ is the instantaneous atomic density order. 
Expanding the wave function in momentum space as
$\psi(\theta,\tau)= \frac{1}{\sqrt{2\pi}}
    \sum_j C_j(\tau)e^{ij\theta}$,
the density modulation can equivalently be written as
$\Theta(\tau) =-\sum_j {\rm Re}[C_{j+1}^*(\tau)C_j(\tau)]$.

Experimentally, the recoil frequency, cavity detuning, and cavity loss can all be on the order of a few kHz~\cite{Klinder2015,Ferri2021}. Choosing the kick period $T$ on the order of milliseconds therefore allows the effective Planck constant $\hbar_s$ to reach the values considered in the main text. The transverse pump can be applied in short pulses with durations on the order of tens of microseconds, which is sufficient to approximate the delta-kick limit~\cite{Cao2022a,SeeToh2022,Guo2025}. For the typical superradiant driving strength $\eta\sim 10$ considered here, the required effective two-photon Raman scattering rate is on the order of a few hundred kHz, which is accessible in current cavity-BEC platforms~\cite{Baumann2010c,Klinder2015,Ferri2021}, where collective enhancement from atom numbers of order $10^5$ can be achieved.


\subsection{B. Critical superradiant instability analysis}
\label{sec:appC_closed_loop_boundaries}

\textbf{B1. Dark states below threshold.} For the dark states below the superradiant threshold, we have 
\begin{eqnarray}
 \psi(\theta,\tau)&=&\psi_s(\theta,\tau)\\
 \alpha(\tau)&=&0
\end{eqnarray}
where the atomic wavefunction satisfies
\begin{eqnarray}
\Theta(n_-)=-\int_0^{2\pi}|\psi_s(\theta,n_-)|^2\cos(\theta)=0 \quad {\rm for }  \quad \forall n 
\end{eqnarray}

(i) When $\hbar_s=\pi$, the atomic kinetic energy for momentum states $j$ reads
$\frac{\hbar_sj^2}{2}$, after folding the free-evolution phase 
into the Floquet Brillouin zone~\cite{Izrailev1980,Lepers2008},
momentum states with even and odd indices acquire only two distinct kinetic energy $0$ and $\pi/2$. 
\begin{enumerate}
    \item For red-detuned driving $\Delta_c<0$, the low energy uniform condensate is stable below threshold $\psi_s(\theta,\tau)=\frac{1}{\sqrt{2\pi}}$. Therefore, $\Theta(\tau)=0$ for arbitrary $\tau$.
    \item However, for blue-detuned driving $\Delta_c>0$, the low energy uniform condensate with $j=0$ is unstable below threshold, the two-photon scattering from driving photon to cavity photon will transfer energy to the atoms and excite them to the high-energy state with odd $j$. Cavity loss sustains such two-photon scattering until the system settles into a new stable dark state.
We find such a stable dark state is a Bessel state 
\begin{eqnarray}
    \psi_s(\theta,n_-)=\frac{e^{-iA\cos\theta}}{\sqrt{2\pi}}=\sum_{j=-\infty}^{\infty} \frac{1}{\sqrt{2\pi}}C^s_j e^{i j \theta},
\end{eqnarray}
with Jacobi-Anger expansion coefficients $C^s_j=(-i)^j J_j(A)$. This state has no density modulation and thus satisfies $\Theta(n_-)=0$. However, since there is no kick, after a free evolution to the next kick, one has
\begin{eqnarray}
    \psi_s(\theta,n_-+1)=e^{i\pi\partial_\theta^2/2}
\frac{e^{-iA\cos\theta}}{\sqrt{2\pi}}=\sum_{j=-\infty}^{\infty} \frac{1}{\sqrt{2\pi}}(-i)^{j^2}(-i)^j J_j(A) \, e^{i j \theta}.
\end{eqnarray}
We have
\begin{eqnarray}
    \Theta(n_-+1) &=&-\sum_j {\rm Re}[C_{j+1}^*(n_-+1)C_j(n_-+1)]\nonumber\\
    &=&-\sum_j {\rm Re}[J_{j+1}(A)J_j(A) i^{(j+1)^2-j^2}i]=\sum_j {\rm Re}[J_{j+1}(A)J_j(A) (-1)^{j}]=J_1(2A)
\end{eqnarray}
To ensure $\Theta(n_-+1)=0$, as we stated in the main text, $A$ is determined by
\begin{eqnarray}
    J_1(2A)=0.
\end{eqnarray}
After excluding the trivial uniform solution \(A=0\), the lowest nonzero Bessel dark state is fixed by the first positive zero of \(J_1\). 
\end{enumerate}

(ii) When $\hbar_s=6.8$, 
the energy difference between neighboring momentum states
is $\delta E_{j+1,j}=\hbar_s(j+1)^2/2-\hbar_sj^2/2=\frac{\hbar_s(2j+1)}{2}$. We consider $j>0$,
after folding into the Floquet Brillouin zone $\delta E^F_{j+1,j}=\delta E_{j+1,j}-2l\pi\in[-\pi,\pi]$,
we find that $\delta E^F_{j+1,j}<0$ for $j<6$ and $\delta E^F_{j+1,j}>0$ for $j=6$. 
This indicates that $j=0$ is a local energy  maximum while $j=\pm6$ are local energy minima.
\begin{enumerate}
    \item For red-detuned driving $\Delta_c<0$, the system is driven to the low energy state at $j=\pm6$. Due to the symmetry, we have $\psi_s(\theta,\tau)=\frac{\cos(6\theta)}{\sqrt{\pi}}$ and  $\Theta(\tau)=0$ for arbitrary $\tau$.
    \item For blue-detuned driving $\Delta_c>0$, the  high-energy uniform condensate is stable below threshold, and we have $\psi_s(\theta,\tau)=\frac{1}{\sqrt{2\pi}}$ and $\Theta(\tau)=0$ for arbitrary $\tau$.
 
\end{enumerate}

\textbf{B2. Instability analysis.} The stroboscopic evolution of the kicked atom-cavity system, from immediately before the \(n\)-th kick to immediately before the $(n+1)$-th kick, is governed by
\begin{equation}
\begin{aligned}
\psi(\theta,n_-+1)
&=
e^{i\hbar_s\partial_\theta^2/2}
e^{-iK(n_-)\cos\theta}
\psi(\theta,n_-),\\
\alpha(n_-+1)
&=
e^{i\Delta_c-\kappa}
\left[
\alpha(n_-)+i\eta\Theta(n_-)
\right].
\end{aligned}
\label{eq:appC_stroboscopic_map}
\end{equation}
We consider small fluctuations around the normal-state solution
$\delta\alpha(n_-),
\delta\Theta(n_-)$.
$\delta\Theta(n_-)$ is induced by the wavefunction fluctuation $\psi(\theta,n_-)=\psi_s(\theta,n_-)+\delta\psi(\theta,n_-)$,
and the kicking fluctuation is given by
$\delta K(n_-)=2\eta\mathrm{Re}[\delta\alpha(n_-)]$. Note that
the normal-state solution satisfies
$\alpha(n_-)=\Theta(n_-)=0$.
Linearizing Floquet map yields a Floquet stability matrix of the fluctuations over one driving period,
and we seek Floquet eigenmodes satisfying
\begin{eqnarray}
    [\delta\alpha(n_-+1),\delta\Theta(n_-+1)]
=\lambda
[\delta\alpha(n_-),\delta\Theta(n_-)],
\label{eq:Floquet_eigS}
\end{eqnarray}
where $\lambda$ is the Floquet multiplier~\cite{1998}. Since only the real part of the cavity field enters the feedback loop and the density order parameter $\Theta$ is real, the relevant instability channels are characterized by real Floquet multipliers. The normal phase is stable when $|\lambda|<1$, while $|\lambda|>1$ signals an instability toward a superradiant state. The transition boundary is therefore determined by $\lambda=\pm1$, corresponding to the symmetric and antisymmetric instability channels.

From Eq.~(\ref{eq:appC_stroboscopic_map}), one obtains
\begin{equation}
\delta\alpha(n_-)
=
\eta\chi_c(\Delta_c,\kappa,\lambda)\delta\Theta(n_-),
\qquad {\rm with} \quad
\chi_c(\Delta_c,\kappa,\lambda)
=
\frac{
i e^{i\Delta_c-\kappa}
}{
\lambda-e^{i\Delta_c-\kappa}
}.
\label{eq:appC_chic}
\end{equation}
The atomic response to the fluctuation is more complicated,
\begin{eqnarray}
    \psi(\theta,n_+)=
e^{-i\delta K(n_-)\cos\theta}
\psi_s(\theta,n_-)+\delta \psi_s(\theta,n_-)\simeq \psi_s(\theta,n_-)-i\psi_s(\theta,n_-)\delta K(n_-)\cos\theta + \delta \psi_s(\theta,n_-).
\end{eqnarray}
In the momentum space, we have
\begin{eqnarray}
    \psi(\theta,n_+)=\frac{1}{\sqrt{2\pi}}\sum_j\left[C^s_j(n_-)+\delta C_j(n_-)- i\delta K(n_-) \frac{C_{j+1}^s(n_-)+C_{j-1}^s(n_-)}{2}\right]e^{ij\theta}
\end{eqnarray}
Therefore
\begin{eqnarray}
    \psi(\theta,n_-+1)=\frac{1}{\sqrt{2\pi}}\sum_j\left[C^s_j(n_-)+\delta C_j(n_-)- i\delta K(n_-) \frac{C_{j+1}^s(n_-)+C_{j-1}^s(n_-)}{2}\right]e^{ij\theta}e^{-i\frac{\hbar_s}{2}j^2},
\end{eqnarray}
we can obtain (the coordinate $n_-$ is dropped for simplicity)
\begin{eqnarray}
\lambda\delta\Theta&=&-\lambda {\rm Re}\sum_j
\bigg[
(C_j^s)^*\,\delta C_{j+1}
+
(\delta C_j)^*\,C_{j+1}^s \bigg] \nonumber\\
&=&\delta \Theta(n_-+1)\nonumber \\
&=&
-{\rm Re}\sum_j
e^{-i\hbar_s\left(j+\frac12\right)}
\Bigg[
(C_j^s)^*\,\delta C_{j+1}
+
(\delta C_j)^*\,C_{j+1}^s \nonumber \\
&-&
\frac{i\delta K}{2}
\bigg(
(C_j^s)^*
\left(
C_{j+2}^s+C_j^s
\right)
-
\left[
(C_{j+1}^s)^*
+
(C_{j-1}^s)^*
\right]
C_{j+1}^s
\bigg)
\Bigg].
\label{eq:C_j_Supp}
\end{eqnarray}

Firstly, we consider the plane wave dark state $C_j=\delta_{j,j_0}$.
Then the solution to Eq.~(\ref{eq:C_j_Supp}) is
\begin{eqnarray}
\lambda \delta C_{j_0+1}&=&
e^{-i\hbar_s\left(j_0+\frac12\right)}(\delta C_{j_0+1}
-\frac{i\delta K}{2})\\
\lambda \delta C_{j_0-1}&=&
e^{i\hbar_s\left(j_0-\frac12\right)}(\delta C_{j_0-1}
-\frac{i\delta K}{2})
\end{eqnarray}
After solving $\delta C_{j_0\pm1}$ as functions of $\delta K$, we obtain
\begin{eqnarray}
    \Theta=-{\rm Re} (\delta C_{j_0+1}+\delta C_{j_0-1})=-\frac{\delta K}{4}\left[
\frac{\sin\Omega_+}{\cos\Omega_+-\lambda}
+
\frac{\sin\Omega_-}{\cos\Omega_--\lambda}
\right]=-\frac{\eta{\rm Re} [\alpha]}{2}\left[
\frac{\sin\Omega_+}{\cos\Omega_+-\lambda}
+
\frac{\sin\Omega_-}{\cos\Omega_--\lambda}
\right]
\end{eqnarray}
with 
\begin{equation}
\Omega_+
=
{\hbar_s}(j_0+\frac12),
\qquad
\Omega_-
=-
{\hbar_s}(j_0-\frac12).
\label{eq:appC_Omega_pm}
\end{equation}
Finally, we obtain
\begin{eqnarray}
   \chi_a(\psi_s,\hbar_s,\lambda)=-\frac{1}{2}\left[
\frac{\sin\Omega_+}{\cos\Omega_+-\lambda}
+
\frac{\sin\Omega_-}{\cos\Omega_--\lambda}
\right]
\end{eqnarray}
The response for $C_j^s=\delta_{j,-j_0}$ is the same as above, therefore
$\psi_s\sim\cos(j_0\theta)$ has the same $\chi_a$ if $j_0>1$. Now, it is straightforward to calculate the critical instability for $j_0=0$
and $j_0=\pm6$ for arbitrary $\hbar_s$.  
That is, from
$\delta\alpha(n_-)=\eta\cdot \chi_c(\Delta_c,\kappa,\lambda)\cdot\delta\Theta(n_-)$ and $\delta\Theta(n_-)=\eta\cdot \chi_a(\psi_s,\hbar_s,\lambda)\cdot\mathrm{Re}[\delta\alpha(n_-)]$, 
we obtain $\delta\Theta(n_-)=\eta^2\chi_a{\rm Re}[\chi_c]\delta\Theta(n_-)$, and thus
the critical pump strength is determined as 
$\eta_c(\lambda)=1/\sqrt{\chi_a\cdot\mathrm{Re}[\chi_c]}$ with $\lambda=\pm1$.


Secondly, we consider the Bessel dark state
$\psi_s(\theta,n_-)=\frac{1}{\sqrt{2\pi}}e^{-iA\cos(\theta)}$.
As we discussed above,
during free evolution,
the Bessel coefficients acquire the dynamical phase factors that enter the density order. At $\tau=n_-+1$, 
the stroboscopic density order reads
\begin{eqnarray}
    \Theta(n_-+1) &=&-\sum_j {\rm Re}[C_{j+1}^*(\tau)C_j(\tau)]\nonumber\\
    &=&J_1(2A)
\end{eqnarray}
with $2A$
the first positive zero of \(J_1\). 
Following the same procedure as in Eq.~(\ref{eq:C_j_Supp}), we can derive that
\begin{eqnarray}
\delta \Theta(n_-+2)&=&-\delta \Theta(n_-)
-{\rm Re}\sum_j
e^{-i\pi\left(j+\frac12\right)}
\Bigg[\frac{i\delta K(n_-+1)}{2}
\bigg(
(C_j^s)^*
\left(
C_{j+2}^s+C_j^s
\right)
-
\left[
(C_{j+1}^s)^*
+
(C_{j-1}^s)^*
\right]
C_{j+1}^s
\bigg)
\Bigg] \nonumber\\
&=&-\delta \Theta(n_-)
-\frac{\delta K(n_-+1)}{2}{\rm Re}\sum_j
(-1)^j\bigg(
(C_j^s)^*
\left(
C_{j+2}^s+C_j^s
\right)
-
\left[
(C_{j+1}^s)^*
+
(C_{j-1}^s)^*
\right]
C_{j+1}^s
\bigg).
\label{eq:C_j_Supp2}
\end{eqnarray}
Based on $C^s_j=(-i)^j J_j(A)$ and the properties of Bessel functions,
we can derive that
\begin{eqnarray}
\delta \Theta(n_-+2)=\lambda^2\delta \Theta(n_-)&=&-\delta \Theta(n_-)
+{\delta K(n_-+1)}\frac{dJ_1(2A)}{dA}\\
&=&-\delta \Theta(n_-)
+\lambda{\delta K(n_-)}\frac{dJ_1(2A)}{dA}
\label{eq:C_j_Supp3}
\end{eqnarray}
which leads to
$2\delta \Theta=
\lambda{\delta K}\frac{dJ_1(2A)}{dA}=2\lambda\delta K J_0(2A)=4\lambda\eta{\rm Re}[\alpha]J_0(2A)$
and thus
\begin{eqnarray}
    \chi_a=2\lambda J_0(2A).
\end{eqnarray}

\textbf{B3. Superradiant phase boundary.}
The critical pump strength is determined as 
\begin{eqnarray}
\eta_c(\lambda)=1/\sqrt{\chi_a\cdot\mathrm{Re}[\chi_c]}\quad \text{with}\quad \lambda=\pm1.
\end{eqnarray}
Using
\begin{equation}
\mathrm{Re}\,\chi_c(\Delta_c,\kappa,\lambda)
=
-\lambda
\frac{
e^{-\kappa}\sin\Delta_c
}{
1-2\lambda e^{-\kappa}\cos\Delta_c+e^{-2\kappa}
},
\label{eq:appC_chic_real_unified}
\end{equation}
we obtain
\begin{equation}
\eta_c(\psi_s,\hbar_s,\lambda;\Delta_c)
=
\sqrt{
\frac{
1-2\lambda e^{-\kappa}\cos\Delta_c+e^{-2\kappa}
}{
-\lambda e^{-\kappa}\sin\Delta_c\,
\chi_a(\psi_s,\hbar_s,\lambda)
}
},
\qquad
\lambda=\pm1 .
\label{eq:appC_eta_candidate_unified}
\end{equation}

\begin{figure}[htbp]
\centering
\includegraphics[width=0.8\linewidth]{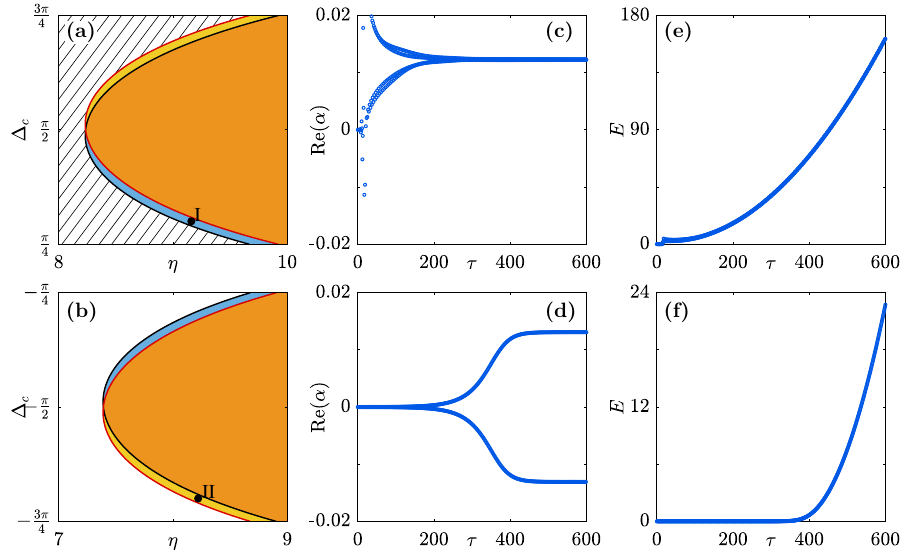}

\caption{The critical superradiant thresholds of symmetric and antisymmetric instability channels for resonant kicking $\hbar_s=\pi$. 
(a) and (b) The dynamical phase diagram 
for blue ($\Delta_c>0$) and red ($\Delta_c<0$) detuning. 
The black and red solid curves denote the superradiant thresholds for the symmetric and antisymmetric instability channels.
(c) and (d) Stroboscopic evolution of the cavity field at $\tau=n$, with parameters $(\eta,\Delta_c)=(9.16,0.3\pi)$ and  $(\eta,\Delta_c)=(8.22,-0.7\pi)$ marked by the black solid circles  in (a) and (b), respectively. Both are governed by a single instability channel.
(e) and (f) Time evolution of the atomic kinetic energy $E$ corresponding to (c) and (d), respectively. The cavity loss rate is $\kappa=4$ in all plots.
}
\label{fig:phase_for_sup}
\end{figure}

\begin{enumerate}
    \item For the uniform dark state with resonant kicking $\hbar_s=\pi$ and $\Delta_c<0$, we have
    \begin{equation}
\eta_c
=
\begin{cases}
\sqrt{
\dfrac{
1+2e^{-\kappa}\cos\Delta_c+e^{-2\kappa}
}{
e^{-\kappa}|\sin\Delta_c|
}
},
& -\pi<\Delta_c<-\dfrac{\pi}{2}, \quad \lambda=-1,\\[1.2em]
\sqrt{
\dfrac{
1-2e^{-\kappa}\cos\Delta_c+e^{-2\kappa}
}{
e^{-\kappa}|\sin\Delta_c|
}
},
& -\dfrac{\pi}{2}<\Delta_c<0, \quad \lambda=+1.
\end{cases}
\label{eq:appC_eta_RU}
\end{equation}

\item For the Bessel dark state with resonant kicking $\hbar_s=\pi$ and $\Delta_c>0$,
The  boundary is
\begin{equation}
\eta_c
=
\begin{cases}
\sqrt{
\dfrac{
1-2e^{-\kappa}\cos\Delta_c+e^{-2\kappa}
}{
2e^{-\kappa}\sin\Delta_c\,|J_0(2A)|
}
},
& 0<\Delta_c<\dfrac{\pi}{2},\quad \lambda=+1, \\[1.2em]
\sqrt{
\dfrac{
1+2e^{-\kappa}\cos\Delta_c+e^{-2\kappa}
}{
2e^{-\kappa}\sin\Delta_c\,|J_0(2A)|
}
},
& \dfrac{\pi}{2}<\Delta_c<\pi \quad \lambda=-1.
\end{cases}
\label{eq:appC_eta_BD}
\end{equation}

\item 
For the uniform dark state with incommensurate kicking $\hbar_s=6.8$  and $\Delta_c>0$, the instability is dominated by the antisymmetric channel \(\lambda=-1\). That is
\begin{equation}
\eta_c
=
\sqrt{
\frac{
1+2e^{-\kappa}\cos\Delta_c+e^{-2\kappa}
}{
e^{-\kappa}\sin\Delta_c\,
|\chi_a(\psi_0,6.8,-1)|
}
},
\qquad
0<\Delta_c<\pi, \quad \lambda=-1.
\label{eq:appC_eta_IU}
\end{equation}
Here, $\psi_0=\frac{1}{\sqrt{2\pi}}$ is the uniform dark state.

\item 
For the finite-momentum-selected dark state with incommensurate kicking $\hbar_s=6.8$ and $\Delta_c<0$, the instability is dominated by the symmetric channel \(\lambda=+1\). That is
\begin{equation}
\eta_c
=
\sqrt{
\frac{
1-2e^{-\kappa}\cos\Delta_c+e^{-2\kappa}
}{
e^{-\kappa}|\sin\Delta_c|\,
|\chi_a(\psi_6,6.8,+1)|
}
},
\qquad
-\pi<\Delta_c<0, \quad \lambda=+1.
\label{eq:appC_eta_FM}
\end{equation}
Here, $\psi_6=\frac{\cos(6\theta)}{\sqrt{\pi}}$ is the momentum-selected dark state.

\end{enumerate}
Equations~\eqref{eq:appC_eta_RU}--\eqref{eq:appC_eta_FM} show that all four theoretical boundaries are determined by the same feedback criterion $\eta_c(\lambda)=1/\sqrt{\chi_a\cdot\mathrm{Re}[\chi_c]}$. The different dark states enter only through the atomic response \(\chi_a(\psi_s,\hbar_s,\lambda)\)

Our analytic phase boundaries are in good agreement with the numerical simulations. 
For resonant kicking with $\hbar_s=\pi$, the critical pump strengths of the two instability channels are different in general, although they may be close to each other, as shown in Figs.~\ref{fig:phase_for_sup}(a) and ~\ref{fig:phase_for_sup}(b). Particularly, the critical boundaries of two channels coincide at $\Delta_c=\pm \frac{\pi}{2}$, leading to balanced response and feedback-induced antiresonance.  
Away from $\Delta_c=\pm \frac{\pi}{2}$, the superradiance exhibits mixed responses of the two channels deep in the resonant phase; while it may be governed by a single instability channel (i.e., $K_2=\pm K_1$) near the superradiant threshold, which also leads to resonance, as confirmed by our numerical simulation as shown in Figs.~\ref{fig:phase_for_sup}(c)-(f).

\subsection{C. Floquet transfer-matrix analysis of the atomic dynamics}
For quantum-resonance condition with $\hbar_s=4\pi\frac{p}{q}$, after folding to the Floquet Brillouin zone, the kinetic energy has a period of $q$ in momentum space. 
Therefore, the kick operation is diagonal in $\theta$ space, while the free evolution only couple $\psi(\theta)$ with
$\psi(\theta+j\frac{2\pi}{q})$ with integer $j\in[0,q-1]$.
We can rewrite the wavefunction as~\cite{Izrailev1980}
\begin{eqnarray}
    \psi(\theta)=\psi_j(\vartheta)
\end{eqnarray}
with 
\begin{equation}
\theta=\vartheta+\frac{2\pi j}{q},
\qquad j=0,1,\cdots,q-1 ,
\end{equation}
and the reduced angular variable $\vartheta\in\left[0,\frac{2\pi}{q}\right]$.
The kick operation is simply
\begin{eqnarray}
    \psi_j(\vartheta,n_+)&=& D_j(K,\vartheta)\psi_j(\vartheta,{n_-}) 
\end{eqnarray}
with $D_j(K,\vartheta) =e^{-iK({n_-})\cos(\vartheta+2j\pi/q)}$. While the free evolution only couples $\psi_j(\vartheta)$ with $\psi_{j'}(\vartheta)$, the coupling coefficient is $\vartheta$ independent which reads
\begin{equation}
\Gamma_{jj'}
=
\frac1q
\sum_{s=0}^{q-1}
\exp\left(-i\frac{\hbar_s}{2}s^2\right)
\exp\left[i s\frac{2\pi}{q}(j-j')\right].
\label{eq:B26_kernel}
\end{equation}
The Floquet evolution during one kick period reads
\begin{eqnarray}
    [\psi_0(\vartheta,n_-+1),\cdots,\psi_{q-1}(\vartheta,n_-+1)]^T=\Gamma D [\psi_0(\vartheta,n_-),\cdots,\psi_{q-1}(\vartheta,n_-)]^T
\end{eqnarray}
with $\Gamma$ and $D$ the evolution given above.
The unitary Floquet transfer matrix is defined as
\begin{equation}
\mathcal S(\vartheta)=\Gamma D.
\label{eq:B10}
\end{equation}
If we assume its eigenvalues as
$\varepsilon_j(\vartheta)=e^{i\phi_j(\vartheta)}$,
this eigenmode leads to an averaged momentum after $n$ kicks as
$p(n)\sim-i(\varepsilon_j^*)^n\partial_\vartheta (\varepsilon_j^n)=n\partial \phi_j(\vartheta)$,
which leads to
\begin{equation}
E(n)\propto n^2 [\partial \phi_j(\vartheta)]^2.
\label{eq:B14}
\end{equation}
When an eigenmode satisfies $\partial_\vartheta\phi_j(\vartheta)\neq0$
and the initial state has a nonzero projection onto this mode, we would expect resonant ballistic transport.
Conversely, if all relevant eigenvalues are independent of $\vartheta$, the coherent momentum-accumulation term proportional to \(n\) is absent, and the leading \(n^2\) contribution to the energy does not appear.

The above transfer matrix analysis can be generalized to the double kick sequences.
For our case with $\hbar_s=\pi$ (i.e., $p=1,q=4$), 
we define
\begin{equation}
\begin{pmatrix}
{\psi_0}(\vartheta)\\
{\psi_1}(\vartheta)\\
{\psi_2}(\vartheta)\\
{\psi_3}(\vartheta)
\end{pmatrix}=
\begin{pmatrix}
\psi(\vartheta)\\
\psi(\vartheta+\pi/2)\\
\psi(\vartheta+\pi)\\
\psi(\vartheta+3\pi/2)
\end{pmatrix},
\label{eq:B21}
\end{equation}
and the two-kick transfer matrix reads
\begin{equation}
S_{\rm eff}(\vartheta)
=
\Gamma D(K_2,\vartheta)\Gamma D(K_1,\vartheta).
\label{eq:B30}
\end{equation}
Substitution of the explicit matrices gives the full two-step matrix
\begin{equation}
S_{\rm eff}(\vartheta)
=
\begin{pmatrix}
-\sin X_2 e^{-iX_1} & 0 & \cos X_2 e^{iX_1} & 0\\
0 & \sin Y_2 e^{iY_1} & 0 & \cos Y_2 e^{-iY_1}\\
\cos X_2 e^{-iX_1} & 0 & \sin X_2 e^{iX_1} & 0\\
0 & \cos Y_2 e^{iY_1} & 0 & -\sin Y_2 e^{-iY_1}
\end{pmatrix}.
\label{eq:B39}
\end{equation}
with $
X_i=K_i\cos\vartheta$, $Y_i=K_i\sin\vartheta$.
The eigenvalues are
\begin{equation}
\varepsilon_{02}^{\pm}(\vartheta)
=
i\sin X_1\sin X_2
\pm
\sqrt{1-\sin^2X_1\sin^2X_2},
\label{eq:B42}
\end{equation}
and
\begin{equation}
\varepsilon_{13}^{\pm}(\vartheta)
=
i\sin Y_1\sin Y_2
\pm
\sqrt{1-\sin^2Y_1\sin^2Y_2}.
\label{eq:B43}
\end{equation}
For finite kick strengths $K_1\neq0$ and $K_2\neq0$,
the eigenvalues in Eqs.~\eqref{eq:B42} and \eqref{eq:B43} depend on the continuous variable \(\vartheta\) and lead to
resonant quadratic energy growth.

Particularly, when the second kick is completely suppressed $K_2=0$, we have $X_2=Y_2=0$ and 
$\varepsilon_{02}^{\pm}=\pm1,
\varepsilon_{13}^{\pm}=\pm1$
independent of $\vartheta$.
Moreover, \(K_2=0\) not only removes the quadratic growth but also yields an exact four-kick recurrence with
\begin{equation}
S_{\rm eff}^2=I.
\label{eq:B50}
\end{equation}
This is precisely the antiresonance condition, here it does not arise from breaking the resonance condition \(\hbar_s=\pi\), but comes from the feedback-generated condition \(K_2=0\) which removes the \(\vartheta\) dependence of the two-step Floquet eigenvalues and further produces an exact recurrence.

\begin{figure}[bp]
\centering
\includegraphics[width=0.55\linewidth]{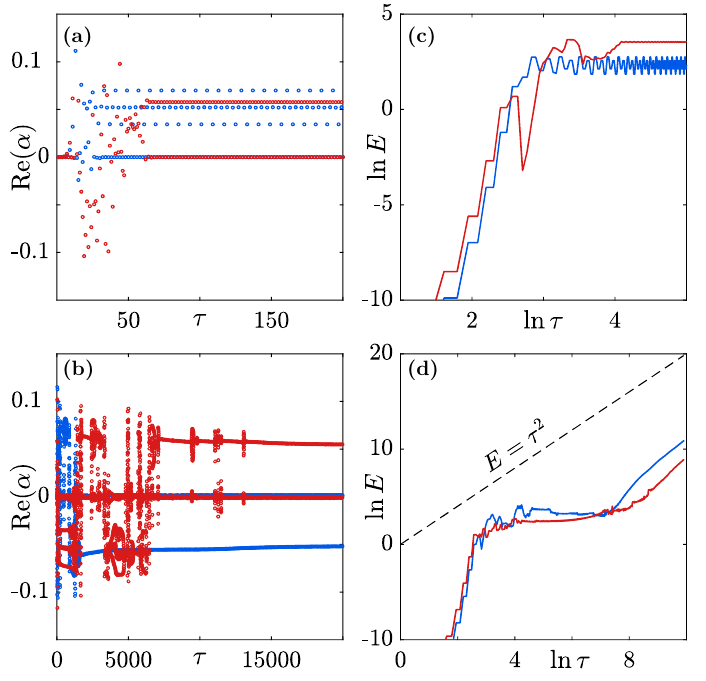}
\caption{Dynamics in the strong driving regime for resonant kicking $\hbar_s=\pi$.
(a) and (b) Stroboscopic evolution of the cavity field for antiresonant and resonant dynamics, with parameters $(\eta,\Delta_c)=(12,-0.5\pi)$ and $(\eta,\Delta_c)=(12,-0.6\pi)$, respectively. (c) and (d)
The evolution of atomic kinetic energy $E$ 
corresponding to (a) and (b), respectively.
The blue and red dots/lines denote different initial cavity field fluctuations.
The black dashed line in (d) indicates $E=\tau^2$.
The cavity loss is $\kappa=4$ in all plots.
}
\label{fig:unstable_dynamic_S}
\end{figure}

\subsection{D. Properties of the superradiant phases}

Firstly, we discuss the feedback-induced antiresonant dynamics with $\hbar_s=\pi$.
Generally, the condensate returns to its initial state only after two  effective kick cycles. Starting from $\tau=n_-$, we have
\begin{equation}
\begin{aligned}
e^{-iA_1\cos\theta} &\xrightarrow{\text{kick}} & e^{-iA_2\cos\theta} 
  &\xrightarrow{\text{free}}& e^{iA_2\cos\theta} 
  &\xrightarrow{\text{kick}}& e^{iA_1\cos\theta} 
  &\xrightarrow{\text{free}}& e^{-iA_1\cos\theta}.
\end{aligned}
\end{equation}
The kick occurs only at $\tau=n_-,n_-+2,n_-+4,\cdots$, it generates the phase $e^{-iK(n_-)\cos(\theta)}$ with $A_1+K(n_-)=A_2$ such that it drives the atom from $e^{-iA_1\cos\theta}$ to $e^{-iA_2\cos\theta}$ and from $e^{iA_2\cos\theta}$ to $e^{iA_1\cos\theta}$, the free evolution of $2T$ between two kicks will reverse the sign of $A_i$. The atomic dynamics is period-quadrupled.
Also note that, at $\tau=n_-+1,n_-+3,\cdots$, the density order $\Theta$ is nonzero and pumps the cavity field and restores its real component for the following kick. That is, $\Theta(n_-+1)=J_1(2A_2)=J_1(-2A_1)\neq 0$. We want to point out that, when $A_1=-A_2$, the atom dynamics 
reduces to period-doubled.
If we start from the Bessel state $e^{-iA\cos(\theta)}$ for $J_1(2A)=0$ and $\Delta_c=\frac{\pi}{2}$, then the small deviation from this state leads to $A_1\lesssim A\lesssim A_2$ satisfying $J_1(2A_2)=J_1(-2A_1)$ but $A_1\neq-A_2$.
However, when we start from uniform state with $A=0$ for red detuning $\Delta_c=-\frac{\pi}{2}$, the small deviation from the uniform state leads to $A_1\lesssim 0\lesssim A_2$ satisfying $J_1(2A_2)=J_1(-2A_1)$; the solution reads $A_1=-A_2$ and thus there is 
a period-doubled antiresonant regime ($\eta<\eta_\star$), in which the condensate recovers after one effective kick cycle.
In the strong pump regime with $\eta>\eta_\star$, the amplitude becomes $A_2>A_m$ with $2A_m$ the solution to the first maximum of the Bessel function $J_1$, the period-quadrupled atomic dynamics is restored with solution $A_2>A_m>-A_1$. The critical value $\eta_\star$ can be calculated analytically through the feedback evolution of the self-adaptive antiresonance dynamics: the kick $K=A_2-A_1=2A_m=2\eta_\star|\alpha|e^{-\kappa}$ and the pump $\eta_\star\Theta=\eta_\star J_1(2A_m)=|\alpha|+|\alpha|e^{-2\kappa}$, leading to $\eta_\star=\sqrt{\frac{A_m(e^{\kappa}+e^{-\kappa})}{J_1(2A_m)}}$. For the parameter used in Fig.~3 in the main text, we have $\eta_\star\simeq9.3$. 
The transition from period-doubled to period-quadrupled dynamics can be understood through an instability analysis, which shows that the period-doubled solution is stable in the regime $\eta<\eta_\star$ and unstable  in the regime ($\eta>\eta_\star$) where the period-quadrupled solution becomes the stable one. 
The above discussion has been numerically confirmed by our simulation.

Secondly, we discuss the strong-driving regime for resonant kicking. As stated in the main text, the dynamics exhibits pronounced fluctuations and becomes increasingly sensitive to the initial state in this regime. In particular, the cavity field, and hence the kicking sequence, fluctuates in time, taking the form of $K_1,0,K_1',0,\cdots$ for the antiresonant dynamics and $K_1,K_2,K_1',K_2',\cdots$ for the resonant dynamics, as shown in Fig.~\ref{fig:unstable_dynamic_S}(a) and \ref{fig:unstable_dynamic_S}(b).
Nevertheless, the competition between the symmetric and antisymmetric instability channels continues to organize the Floquet dynamics and sustain robust antiresonant and resonant responses. 
For the antiresonant dynamics, the suppression of every second kick preserves the underlying antiresonant mechanism, and the system remains dynamically localized through destructive interference, as shown in Fig.~\ref{fig:unstable_dynamic_S}(c). Correspondingly, the eigenvalue of the total effective transfer matrix takes the form of $e^{-i\sum_n (-1)^nK(n)\cos(\theta)}$, ensuring periodic evolution and bounded energy growth despite the fluctuating kick amplitudes. For the resonant phase, the finite alternating kick amplitudes preserve the resonant character of the effective Floquet evolution, and the quadratic energy growth $E(\tau)\propto\tau^2$ persists despite the fluctuations, as shown in Fig.~\ref{fig:unstable_dynamic_S}(d).

\begin{figure}[bp]
\centering
\includegraphics[width=0.55\linewidth]{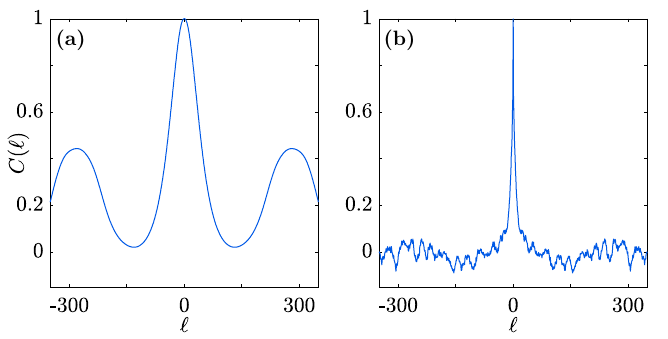}
\caption{Temporal correlation of the self-generated kick strength in the irregular localization phase (a) and subdiffusion phase (b).
$C(\ell)$ is the normalized temporal autocorrelation function. As comparison, white noise has a correlation function $C_{\text{white}}(\ell)=\delta_{\ell,0}$.
The pump strengths are $\eta=8$ in (a) and $\eta=20$ in (b).
Other parameters are: $\hbar_s=6.8$, $\Delta_c=0.5\pi$, and $\kappa=4$. 
}
\label{fig:temp_corr_S}
\end{figure}

Thirdly, we consider the incommensurate regime and analyze the temporal correlations of the feedback-induced kick fluctuations. 
We define the normalized autocorrelation function as
\begin{eqnarray}
    C(\ell)
=
\frac{
\sum_{n}
\left(
x_n-\bar{x}
\right)
\left(
x_{n+|\ell|}
-\bar{x}
\right)
}{\sum_{n}
\left(
x_n-\bar{x}
\right)^2
}
\end{eqnarray}
with $x_n=|K(\tau=2n)|$ and $\bar{x}$ the averaged value of $x_n$, the summation is taken over $n\in[2000,4000]$. Since we are considering the antisymmetric instability superradiant phases, the correlation is evaluated using only the even kicks. The decay of $C(\ell)$ with increasing $|\ell|$ characterizes the temporal
memory of the feedback-induced kick-amplitude fluctuations. A slow decay
indicates that the fluctuations remain strongly correlated over many kick
periods, whereas a rapid decay signals that the kick sequence becomes
increasingly noise-like (white noise has a correlation function $C_{\text{white}}(\ell)=\delta_{\ell,0}$).
As the pump strength increases, the cavity field develops increasingly irregular dynamics, leading to a crossover from the dynamical irregular localization phase to the feedback-induced subdiffusion phase.
As shown in Fig.~\ref{fig:temp_corr_S}(a), weak feedback has small fluctuations that are strongly correlated in time and thus the localization persists. With increasing pump strength, the correlations gradually decay and the kick sequence becomes increasingly noise-like, giving rise to subdiffusive momentum transport with approximately
$E(\tau)\propto\tau^{2/3}$.
Importantly, even in the strong-driving regime the fluctuations are not completely random but retain finite temporal correlations, as shown in Fig.~\ref{fig:temp_corr_S}(b). The correlated nature of the feedback-induced noise distinguishes the present system from a conventional noisy kicked rotor~\cite{Cohen1991d,Ammann1998a,Klappauf1998b,Sarkar2017} and is responsible for the observed anomalous subdiffusive transport.

\begin{figure}[tbp]
\centering
\includegraphics[width=0.8\linewidth]{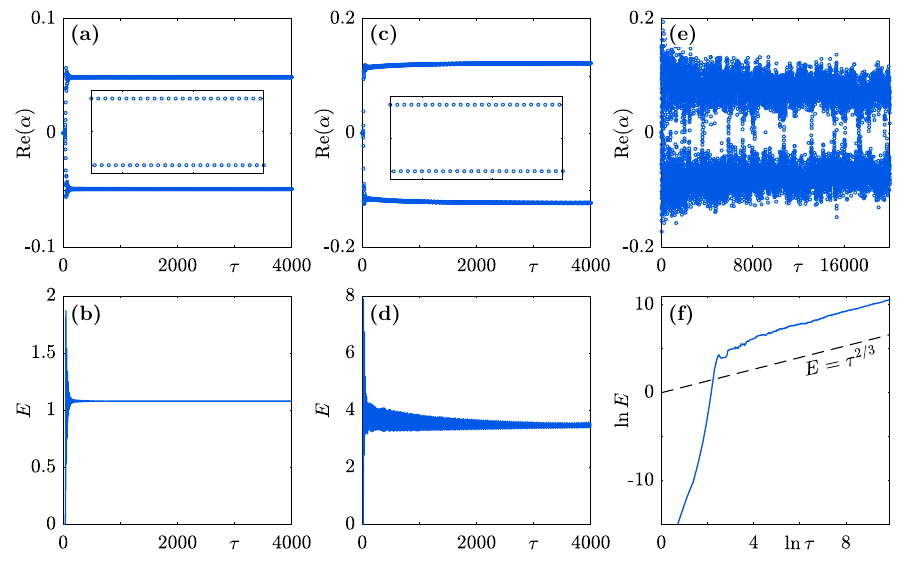}

\caption{Effects of atomic interaction and cavity-induced ac-Stark shift on the dynamical phases under incommensurate kicking $\hbar_s=6.8$. The parameters are $\Delta_c=0.5\pi$, $\kappa=4$, $U=0.1$, and $g_{\rm a}=0.1$. (a) and (b) Time evolution of the cavity field and atomic kinetic energy for the period-doubled localized phase with $\eta=4$. Inset shows a zoom-in around $\tau=4000$. (c) and (d) Same as that in (a) and (b) but for the irregular localized phase with $\eta=8$. Inset shows a zoom-in around $\tau=4000$. (e) and (f) Same as that in (a) and (b) but for the subdiffusion phase with $\eta=20$.
The dynamics is hardly affected by the weak atomic interaction and cavity-induced ac-Stark shift.}
\label{fig:int_acstark_S1}
\end{figure}

\begin{figure}[htbp]
\centering
\includegraphics[width=0.8\linewidth]{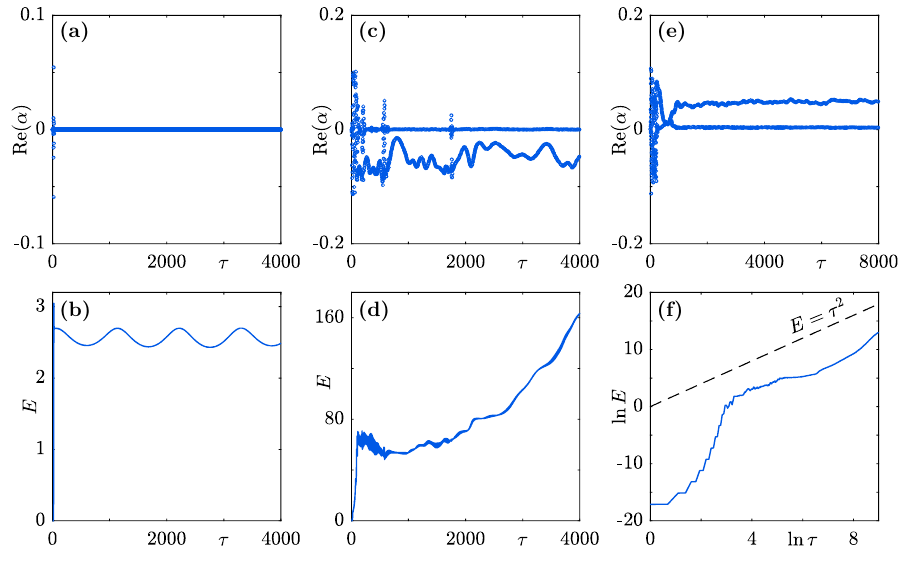}
\caption{Effects of atomic interaction and cavity-induced ac-Stark shift on the dynamical phases under incommensurate kicking $\hbar_s=\pi$. The parameters are $\kappa=4$, $U=0.1$, and $g_{\rm a}=0.1$. (a) and (b) Time evolution of the cavity field and atomic kinetic energy for the Bessel dark phase with $\Delta_c=0.5\pi$ and $\eta=6.5$. (c) and (d) Same as that in (a) and (b) but for the antiresonant phase with $\Delta_c=-0.5\pi$ and $\eta=10$. (e) and (f) Same as that in (a) and (b) but for the resonant phase with $\Delta_c=-0.3\pi$ and  $\eta=11$.
}
\label{fig:int_acstark_S2}
\end{figure} 

\subsection{E. Effects of interaction and cavity-induced ac-Stark shift}
\label{subsec:appA_general_dimensionless_equations}
We now discuss the effects of atomic interaction and the cavity-induced ac-Stark shift, even though they can be tuned to negligibly weak values. For nonzero $g_{\rm a}$ and $U$, 
the kick operation remains unchanged,
the free evolution is modified as
\begin{eqnarray}
    \alpha(n_+)&=&\alpha(n_-)+i\eta\Theta({n_-}), \nonumber \\
    \psi(\theta,n_+)&=& e^{-iK({n_-})\cos\theta}\psi(\theta,{n_-}) 
\end{eqnarray}
with $n_{\pm}$ denoting the time right before and after the kick, respectively. The free cavity evolution between two consecutive kicks is
\begin{equation}
\begin{aligned}
i\partial_\tau\psi(\theta,\tau)
&=
\bigg[-\frac{\hbar_s}{2}\partial_\theta^2
+g_{\rm a}|\psi(\theta,\tau)|^2
+U|\alpha(\tau)|^2\cos^2\theta\bigg]\psi(\theta,\tau).\\
i\partial_\tau\alpha(\tau)
&=
\left[
-\Delta_c-i\kappa
+
U\mathcal B(\tau)
\right]\alpha(\tau)
\end{aligned}
\label{eq:appA_general_atom}
\end{equation}
where \(\mathcal B(\tau)\) is defined as
$\mathcal B(\tau)=\int_0^{2\pi}d\theta\,
    |\psi(\theta,\tau)|^2\cos^2\theta$.

In Fig.~\ref{fig:int_acstark_S1} and Fig.~\ref{fig:int_acstark_S2}, we present numerical simulations of the self-adaptive QKR dynamics in the presence of weak atom-atom interactions and cavity-induced ac-Stark shifts. We find that all dynamical phases identified in the main text remain qualitatively robust against these additional effects, with the notable exception of the feedback-induced antiresonance. Since the antiresonance relies on a delicate balance between the symmetric and antisymmetric instability channels together with coherent destructive interference, it is particularly sensitive to additional nonlinearities and perturbations, underscoring the genuine feedback origin of the destructive interference kicks.
Therefore, the long-time evolution gradually crosses over to delocalized behavior, as shown in Figs.~\ref{fig:int_acstark_S2}(c) and ~\ref{fig:int_acstark_S2}(d). Moreover, atom-atom interactions generate additional phases that induce density modulations in the Bessel dark state, leading to oscillations of the kinetic energy and small but finite kicks that persist even in the long-time limit, as shown in Figs.~\ref{fig:int_acstark_S2}(a) and ~\ref{fig:int_acstark_S2}(b).

We expect that stronger interactions and larger cavity-induced ac-Stark shifts may qualitatively reshape the self-adaptive QKR dynamics and generate additional dark states together with new localization and delocalization phases. Understanding the interplay between feedback, interactions, and nonlinear cavity potentials remains an interesting direction for future investigations.

\end{widetext}

\end{document}